\documentstyle[aps,prd,eqsecnum,epsf]{revtex}

\newcommand{\singlefig}[2]{
\begin{center}
\begin{minipage}{#1}
\epsfxsize=#1
\epsffile{#2}
\end{minipage}
\end{center}}
\newenvironment{figcaption}[2]{
 \vspace{0.3cm}
 \refstepcounter{figure}
 \label{#1}
 \begin{center}
 \begin{minipage}{#2}
 \begingroup \small FIG. \thefigure: }{
 \endgroup
 \end{minipage}
 \end{center}}

\begin{document}

\draft
\title{Power-law inflation with a nonminimally coupled scalar field}
\author{Shinji Tsujikawa \thanks{electronic
address:shinji@gravity.phys.waseda.ac.jp}}
\address{Department of Physics, Waseda University,
3-4-1 Ohkubo, Shinjuku-ku, Tokyo 169-8555, Japan\\[.3em]}
\date{\today}
\maketitle
\begin{abstract}
We consider the dynamics of power-law inflation with a nonminimally coupled 
scalar field $\phi$. It is well known that multiple scalar fields 
with exponential potentials 
$V(\phi)=V_0 {\rm exp}(-\sqrt{16\pi/p m_{\rm pl}^2} \phi)$
lead to an inflationary solution
even if the each scalar field is not capable to sustain inflation.
In this paper, we show that inflation can be assisted
even in the one-field case 
by the effect of nonminimal coupling.
When $\xi$ is positive, since an effective potential which 
arises by a conformal transformation becomes flatter
compared with the case of $\xi=0$ for $\phi>0$, we have 
an inflationary solution even when the universe evolves as
non-inflationary in the minimally coupled case.
For the negative $\xi$, the assisted inflation can take place
when $\phi$ evolves in the region of $\phi<0$ .
\end{abstract}


\baselineskip = 16pt

%
\section{Introduction}                            %

The idea of inflation is one of the most reliable concepts in modern 
cosmology\cite{GS,KT}. It can solve the horizon and flatness 
problem in the standard big bang cosmology, and also 
provides us the seeds of the large scale structure.
The inflationary period proceeds while a scalar field called 
{\it inflaton} slowly evolves along a sufficiently flat potential.

It has been generally considered that only one scalar field determines
the dynamics of inflation even if there are many scalar fields 
present in the inflationary epoch.
However, Liddle, Mazumdar, and Schunck\cite{assisted} 
recently showed that we have
an inflationary solution with exponential 
potentials\cite{LM,Halliwell,BB} in the multi-scalar 
field case even if individual fields do not possess flat potentials
to lead to inflation.
It was demonstrated that scaling solutions for exponential 
potentials with different slopes are the late-time attractors,
and Malik and Wands confirmed this fact
by choosing an appropriate rotation
in field space\cite{MW}. 
Copeland, Mazumdar, and Nunes examined the 
generalized assisted inflation where 
the cross-coupling terms exist between 
scalar fields\cite{CMN}.
Exponential potentials often arise 
in the effective four-dimensional models induced
by Kaluza-Klein theories.
Kanti and Olive\cite{KO} investigated the assisted 
chaotic inflation with multiple scalar fields
in higher-dimensional theories. The dynamics and 
density perturbations in assisted chaotic inflation was studied by 
Kaloper and Liddle\cite{KL}.
Since most models in realistic higher-dimensional 
theories give rise to steep potentials
by which inflation is hard to realize, the assisted mechanism 
by multiple
scalar fields plays an important role for the realization of inflation.

From a viewpoint of quantum field theory in curved spacetime, it is 
natural to consider that the inflaton field $\phi$ couples nonminimally to the
spacetime curvature $R$ with a coupling of $\xi R\phi^2/2$. 
In the new inflation model\cite{newinflation}, the existence
of nonminimal coupling prevents inflation in some cases, because
the flatness of the potential of inflaton is destroyed around $\phi=0$.
In the chaotic inflation model\cite{Chaotic} 
with a nonminimally coupled inflaton field, 
Futamase an Maeda\cite{FM} investigated the constraint 
of the coupling $\xi$ in two potentials of 
$V(\phi)=m^2\phi^2/2$ and $V(\phi)=\lambda\phi^4/4$.
They found that $\xi$ is restricted as 
$|\xi|~\mbox{\raisebox{-1.ex}{$\stackrel
     {\textstyle<}{\textstyle \sim}$}}~10^{-3}$
to lead to sufficient inflation in the massive inflaton model.
On the other hand, the constraint of $\xi$ is absent in the self coupling model
with negative $\xi$, and inflation is supported for
larger values of $|\xi|$.
Fakir and Unruh examined this self-coupling model with a strong 
negative nonminimal coupling $|\xi| \gg 1$\cite{FU}, 
and found that the fine tuning problem of 
the self-coupling $\lambda$ is relaxed by such large 
values of $|\xi|$.
Several authors studied scalar density 
perturbations\cite{MS,Salopek,Kaiser} and tensor gravitational 
waves\cite{KF,HN} during inflation in this model.
In the context of preheating after inflation, we have showed that 
the fluctuation of inflaton can be enhanced nonperturbatively
by nonminimal coupling\cite{preheating}.

In this paper, we consider power-law inflation with a nonminimally coupled 
inflaton field. What we are concerned with is whether
power-law inflation is supported or not 
with the presence of nonminimal coupling.
If the assisted mechanism works even in the one-field case, 
it is expected that this will also occur in the multi-field case.
For a more complicated study of the multi-field case in the future,
in this paper, we clarify in what values of $\xi$ assisted inflation 
is realized by nonminimal coupling in the one-field case.
   
This paper is organized as follows.
In the next section, we explain the model of power-law inflation
with a nonminimally coupled inflaton field $\phi$. 
In Sec.~III, the dynamics of inflation is investigated in 
both cases of $\xi>0$ and $\xi<0$.
We show in what cases the assisted inflation takes place
by the effect of nonminimal coupling. 
We present our conclusions and discussions in the final section.

\section{Basic equations}   

We study a model where an inflaton field $\phi$ is nonminimally 
coupled with a scalar curvature $R$:
\begin{eqnarray}
{\cal L} = \sqrt{-g} \left[ \frac{1}{2\kappa^2}R
   -\frac{1}{2}(\nabla \phi)^2
   -V(\phi)
   -\frac12 \xi R \phi^2
    \right],
\label{B1}
\end{eqnarray}
where $\kappa^{2}/8\pi \equiv G =m_{\rm pl}^{-2} $ is Newton's
gravitational constant, and $\xi$ is a coupling constant.
In this paper, we consider an exponential potential $V(\phi)$ 
which is described by 
\begin{eqnarray}
V(\phi)=V_0~{\rm exp} \left(
-\sqrt{\frac{16\pi}{p}} \frac{\phi}{m_{\rm pl}} \right),
\label{B2}
\end{eqnarray}
where $V_0$ and $p$ denote the energy scale which has 
the dimension of $[{\rm mass}]^4$,
and the power of inflation, respectively.

In the case of $\xi=0$, we have an inflationary solution for
$p>1$,
\begin{eqnarray}
a(t) \propto t^p,
\label{B3}
\end{eqnarray}
where $a(t)$ is the scale factor.
However, inflation does not take place for $p \le 1$ in the case of a single
scalar field.  Liddle et al.\cite{assisted} showed that inflation proceeds for the 
case of multi-scalar fields even if the individual scalar field has the power
less than unity. This cooperative behavior was termed {\it assisted inflation}.
In this paper, we investigate whether nonminimal coupling will assist inflation
or not in the single field case. The multi-scalar field case will 
be discussed elsewhere\cite{comment}.

We find from the Lagrangian $(\ref{B1})$ that the effective gravitational
constant $G_{\rm eff}$ depends on the value of the inflaton field,
\begin{eqnarray}
G_{\rm eff}=\frac{G}{1-\phi^2/\phi^2_c},~~{\rm with}~~
\phi^2_c \equiv \frac{m^2_{\rm pl}}{8\pi\xi}.
\label{B4}
\end{eqnarray}
In order to connect to our present universe, $G_{\rm eff}$
needs to be positive for the case of the positive $\xi$, which yields 
\begin{eqnarray}
|\phi|<\phi_c=\frac{m_{\rm pl}}{\sqrt{8\pi \xi}}.
\label{B5}
\end{eqnarray}
When $\xi$ is negative, such a constraint is absent.

We obtain the following field equations 
from the Lagrangian $(\ref{B1})$,
\begin{eqnarray}
\frac{1-\xi\kappa^2\phi^2}{\kappa^2} G_{\mu\nu}
 &= & (1-2\xi)\nabla_{\mu} \phi \nabla_{\nu} \phi 
-\left(\frac12 -2\xi \right) g_{\mu \nu} (\nabla \phi)^2
 -g_{\mu \nu} V(\phi) +
 2\xi \phi (g_{\mu \nu} \kern1pt\vbox{\hrule height
1.2pt\hbox{\vrule width 1.2pt\hskip 3pt
   \vbox{\vskip 6pt}\hskip 3pt\vrule width 0.6pt}\hrule
height 0.6pt}\kern1pt -\nabla_{\mu}
\nabla_{\nu}) \phi, 
\label{B6}
\end{eqnarray}
\begin{eqnarray}
\kern1pt\vbox{\hrule height
1.2pt\hbox{\vrule width 1.2pt\hskip 3pt
   \vbox{\vskip 6pt}\hskip 3pt\vrule width 0.6pt}\hrule
height 0.6pt}\kern1pt \phi -\xi R \phi -V,_{\phi}=0,
\label{B7}
\end{eqnarray}
where $\kern1pt\vbox{\hrule height
1.2pt\hbox{\vrule width 1.2pt\hskip 3pt
   \vbox{\vskip 6pt}\hskip 3pt\vrule width 0.6pt}\hrule
height 0.6pt}\kern1pt$ and $V,_{\phi}$ are defined as
 $\kern1pt\vbox{\hrule height
1.2pt\hbox{\vrule width 1.2pt\hskip 3pt
   \vbox{\vskip 6pt}\hskip 3pt\vrule width 0.6pt}\hrule
height 0.6pt}\kern1pt \equiv \partial_{\mu}
(\sqrt{-g} g^{\mu \nu} \partial_{\nu})/\sqrt{-g}$,  $V,_{\phi}
\equiv \partial V/\partial \phi$ respectively. 
Although we can analyze the evolution of the system by 
Eqs.~$(\ref{B6})$ and $(\ref{B7})$, it is rather complicated
due to the existence of nonminimal coupling.
It is convenient to transform to the Einstein frame 
by performing a conformal transformation
\begin{eqnarray}
\hat{g}_{\mu \nu}=\Omega^2 g_{\mu \nu},
\label{B8}
\end{eqnarray}
where $\Omega^2 \equiv 1-\xi\kappa^2\phi^2$.
Then we obtain the following equivalent Lagrangian:
\begin{eqnarray}
{\cal L} = \sqrt{-\hat{g}} \left[ \frac{1}{2\kappa^2} \hat{R}
   -\frac{1}{2} F^2 (\hat{\nabla} \phi)^2
   -\hat{V} (\phi) \right],
\label{B9}
\end{eqnarray}
where variables with a caret denote those in the Einstein frame,
and  
\begin{eqnarray}
F^2 \equiv \frac{1-(1-6\xi) \xi\kappa^2 \phi^2 }
{(1-\xi\kappa^2 \phi^2)^2},
\label{B10}
\end{eqnarray}
\begin{eqnarray}
\hat{V} (\phi)
\equiv \frac{V(\phi)}{(1-\xi\kappa^2\phi^2)^2}.
\label{B11}
\end{eqnarray}
Introducing a new scalar field $\Phi$ as
\begin{eqnarray}
\Phi \equiv \int F(\phi) d\phi,
\label{B12}
\end{eqnarray}
the Lagrangian in the new frame is reduced to the canonical form:
\begin{eqnarray}
{\cal L} = \sqrt{-\hat{g}} \left[ \frac{1}{2\kappa^2} \hat{R}
   -\frac{1}{2} (\hat{\nabla} \Phi)^2
   -\hat{V} (\Phi)  \right].
\label{B13}
\end{eqnarray}

In this paper, we adopt the flat Friedmann-Robertson-Walker 
line element  as the background spacetime;
\begin{eqnarray}
ds^2 = -dt^2 + a^2(t) d {\bf x}^2
=\Omega^{-2} (-d\hat{t}^2 + \hat{a}^2(\hat{t}) d {\bf x}^2).
\label{B14}
\end{eqnarray}
Note that $\hat{t}$ and $\hat{a}$ are related with those 
in the original frame as 
\begin{eqnarray}
\hat{t}=\int \Omega dt,~~~\hat{a}=\Omega a.
\label{B15}
\end{eqnarray}

The evolutions of the scale factor and the $\Phi$ field 
in the Einstein frame yield
\begin{eqnarray}
\hat{H}^2 \equiv \left(\frac{\hat{a}_{,\hat{t}}}{\hat{a}}\right)^2
=\frac{\kappa^2}{3}\left[ \frac12 \Phi_{,\hat{t}}^2
+\hat{V}(\Phi) \right],
\label{B16}
\end{eqnarray}
\begin{eqnarray}
\Phi_{,\hat{t} \hat{t}}+3\hat{H}\Phi_{,\hat{t}}+
\hat{V}_{,\Phi}=0,
\label{B17}
\end{eqnarray}
where $,\hat{t}\equiv d/d\hat{t}$ and 
\begin{eqnarray}
\hat{V}_{,\Phi} \equiv \frac{d\hat{V}}{d\Phi}=
\frac{4\xi\kappa^2\phi-
\sqrt{16\pi/p m_{\rm pl}^2}(1-\xi\kappa^2\phi^2)}
{(1-\xi\kappa^2\phi^2)^2 \sqrt{1-(1-6\xi)\xi\kappa^2\phi^2}}
V_0~{\rm exp} \left(
-\sqrt{\frac{16\pi}{p}} \frac{\phi}{m_{\rm pl}} \right).
\label{B40}
\end{eqnarray}
Note that $\phi_{,\hat{t}}$ and $\Phi_{,\hat{t}}$  
are related by Eq.~$(\ref{B12})$ as
\begin{eqnarray}
\phi_{,\hat{t}}=\frac{1-\xi\kappa^2 \phi^2}
{\sqrt{1-(1-6\xi) \xi\kappa^2 \phi^2}} \Phi_{,\hat{t}} .
\label{B18}
\end{eqnarray}
We can know the behavior of the scale factor and 
the inflaton field by Eqs.~$(\ref{B16})$ and  $(\ref{B17})$ 
with an effective potential $(\ref{B11})$ 
in the Einstein frame. Transforming back to the original frame
by making use of the relation $(\ref{B15})$, we can judge
in what cases nonminimal coupling assists inflation. 
In what follows, we will investigate these issues in details.

\section{Assisted inflation with a nonminimally coupled scalar field}   
In this section, we study the dynamics of the system 
in both cases of the positive and negative $\xi$.
Let us first review the case of $\xi=0$ for comparison.
In this case, making use of the slow-roll conditions
$\dot{\phi}^2 \ll V,~\ddot{\phi} \ll 3H\dot{\phi}$,
Eqs.~$(\ref{B16})$ and $(\ref{B17})$ are
approximately written as
\begin{eqnarray}
H^2
&\approx& \frac{\kappa^2}{3} V_0~{\rm exp} \left(
-\sqrt{\frac{16\pi}{p}} \frac{\phi}{m_{\rm pl}} \right), \\
\label{C1} 
3H\dot{\phi} &\approx& -\sqrt{\frac{16\pi}{p}}
\frac{V_0}{m_{\rm pl}} {\rm exp} \left(
-\sqrt{\frac{16\pi}{p}} \frac{\phi}{m_{\rm pl}} \right),
\label{C2}
\end{eqnarray}
where we dropped a caret in the Hubble parameter. 
Then we find that $\phi$ and $H$ evolve as
\begin{eqnarray}
\phi &\approx& \sqrt{\frac{p}{4\pi}}
{\rm log} \left[ \sqrt{\frac{\kappa^2 V_0}{3}} \frac{t}{p}
+\alpha \right] m_{\rm pl}, \\
\label{C3}
H &\approx& \left( \frac{t}{p}+\beta \right)^{-1},
\label{C4}
\end{eqnarray}
where $\alpha$ and $\beta$ are some constants which 
depend on initial values of the $\phi$ field.
We obtain the power-law solution $(\ref{B3})$ 
by integrating Eq.~$(\ref{C4})$.
When $p$ is greater than unity, 
the potential $(\ref{B11})$ is flat enough
to lead to inflation.
Since the exponential potential does not have a local minimum, 
the inflationary phase continues forever.
In realistic models of inflation, we need to consider 
some exit mechanisms from power-law inflation in order to 
lead to the successful reheating, radiation and matter dominant
universe. In the following subsections, 
we consider the effect of nonminimal
coupling in the dynamics of power-law inflation.

\subsection{Case of $\xi>0$}   
In this case, the inflaton field is required to lie 
in the region of $(\ref{B5})$.
An important point with the existence of positive $\xi$ is that an
effective potential $(\ref{B11})$ in the Einstein frame
has a local minimum at
\begin{eqnarray}
\phi_*= \left( \sqrt{\frac{p}{4\pi}+\frac{1}{8\pi\xi}}-
\sqrt{\frac{p}{4\pi}} \right) m_{\rm pl}.
\label{C5}
\end{eqnarray}
Note that $\phi_*$ exists in the range of $0<\phi_*<\phi_c$.
The evolution of the scale factor depends on the initial condition
of the inflaton field. 
When $\phi$ is close to the critical value $\pm \phi_c$,
$\hat{V}$ is approximately written as\cite{FM}
\begin{eqnarray}
\hat{V}(\Phi) \approx A {\rm exp}\left(2\sqrt{\frac23}\kappa \Phi \right),
\label{C6}
\end{eqnarray}
where $A$ is a constant.
This exponential potential has a power-law solution which is 
described by
\begin{eqnarray}
\hat{a} \propto \hat{t}^{3/4}.
\label{C7}
\end{eqnarray}
Going back to the original frame, the scale factor evolves as 
\begin{eqnarray}
a \propto t^{1/2}.
\label{C8}
\end{eqnarray}
This means that we do not have an inflationary solution
when $\phi$ is close to $\pm \phi_c$.

\begin{figure}
\begin{center}
\singlefig{12cm}{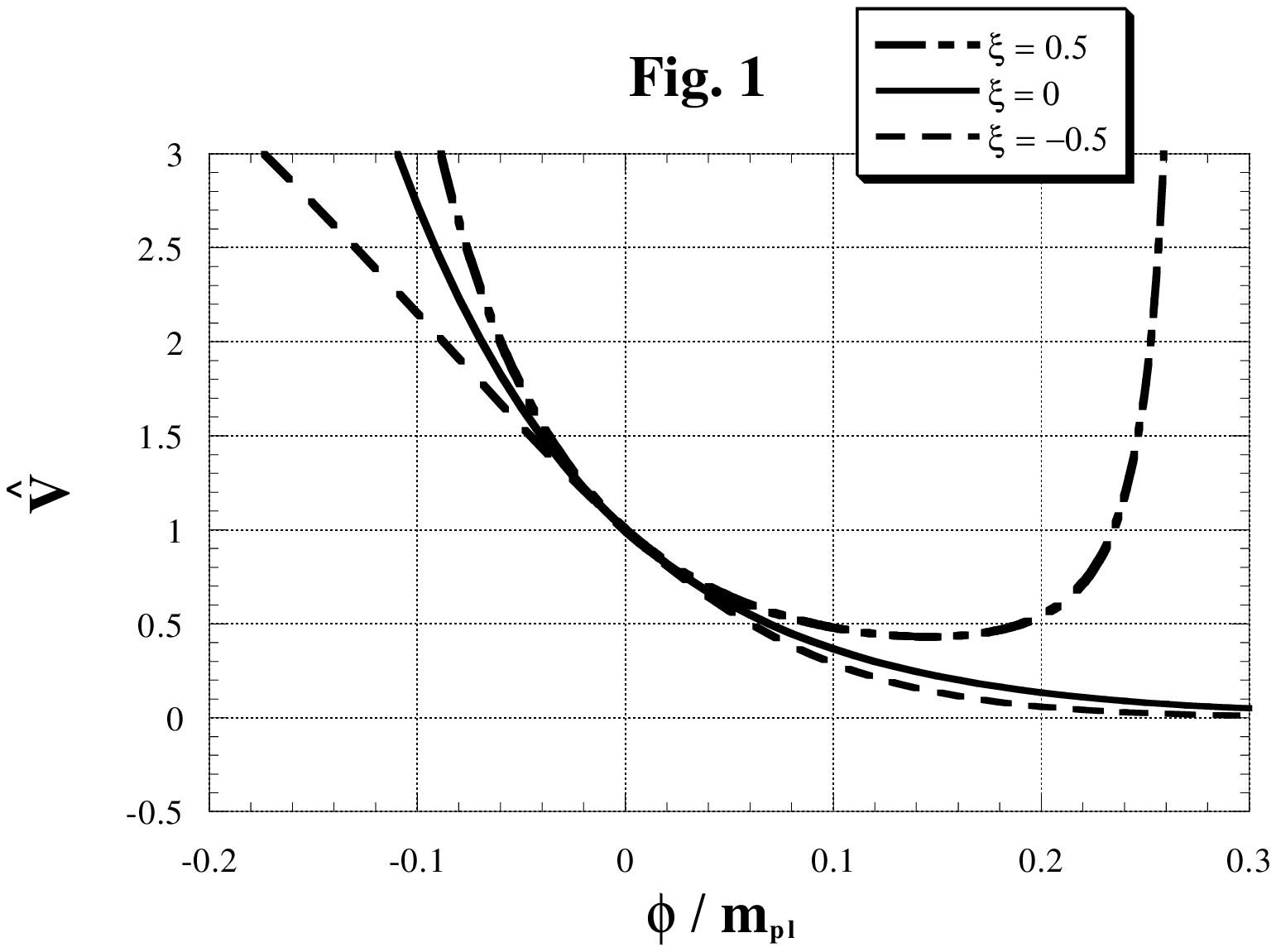}
\begin{figcaption}{Fig1}{12cm}
An effective potential $\hat{V}(\phi)$ in the Einstein frame
in the case of $\xi=0.5, 0, -0.5$ with $p=1/2$
(top, middle, bottom).
When $\xi$ is positive, since the potential $\hat{V}(\phi)$
is flat in the region of $\phi>0$ compared with the $\xi=0$ case,
we can expect assisted inflation to occur in this region.
When $\xi$ is negative, assisted inflation 
can be realized in the region of $\phi<0$.
\end{figcaption}
\end{center}
\end{figure}

Let us consider the case where the initial value 
of inflaton ($=\phi_i$) is in the range of 
$-\phi_c<\phi_i<\phi_*$. Since the slope of the effective
potential $\hat{V}(\phi)$ becomes steeper than 
in the case of $\xi=0$
for the values of $\phi<0$ due to the
existence of $1/(1-\xi\kappa^2\phi^2)^2$
term (See Fig.~1), nonminimal coupling does not 
support inflation when $\phi$ rolls down in the region of $\phi<0$.
However, after inflaton passes through $\phi=0$,
we have a flatter effective potential compared with the 
case of $\xi=0$. 
We can expect that the universe evolves as inflationary
even for the value of $p \le 1$. 
For example,  consider the case of
$p=1/2$ and $\xi=0.05$ with the initial value $\phi_i$
close to $-\phi_c=-0.892 m_{\rm pl}$. 
In Fig.~2, we show the evolution of the scale factor $a$ 
as a function of $t$.
The evolution of the inflaton field $\phi$ is also depicted 
in Fig. 3. Note that these variables are those in the original frame.
We find that inflation does not take place in the region of $\phi<0$.
At the time of $\bar{t} \equiv m_{\rm pl}t \approx 0.1$,
inflaton reaches $\phi=0$. 
After that, the assisted mechanism due to
nonminimal coupling begins to work, and the universe evolves as
inflationary after $\bar{t} \approx 1.5$, which corresponds to 
the value of $\phi \approx 0.5 m_{\rm pl}$.
Even in the case of $0<\phi<0.5 m_{\rm pl}$, the $\xi$ effect 
makes the universe grow more rapidly than in the case of $\xi=0$.
Assisted inflation relevantly occurs in the region of 
$0.5 m_{\rm pl}~\mbox{\raisebox{-1.ex}{$\stackrel
     {\textstyle<}{\textstyle \sim}$}}~\phi~\mbox{\raisebox{-1.ex}{$\stackrel
     {\textstyle<}{\textstyle \sim}$}}~\phi_*=0.714m_{\rm pl}$.
Inflaton slowly evolves in the flat region around $\phi=\phi_*$,
and finally approaches the local minimum.
In the case where inflaton is initially located in the region of 
$0<\phi_i<\phi_*$,
the assisted mechanism works from the beginning. 
After the field reaches the local minimum at $\phi=\phi_*$,
it continues to stay there, and the inflationary phase
does not terminate. Hence we need some exit mechanisms from 
inflation as in the minimally coupled case.

\begin{figure}
\begin{center}
\singlefig{12cm}{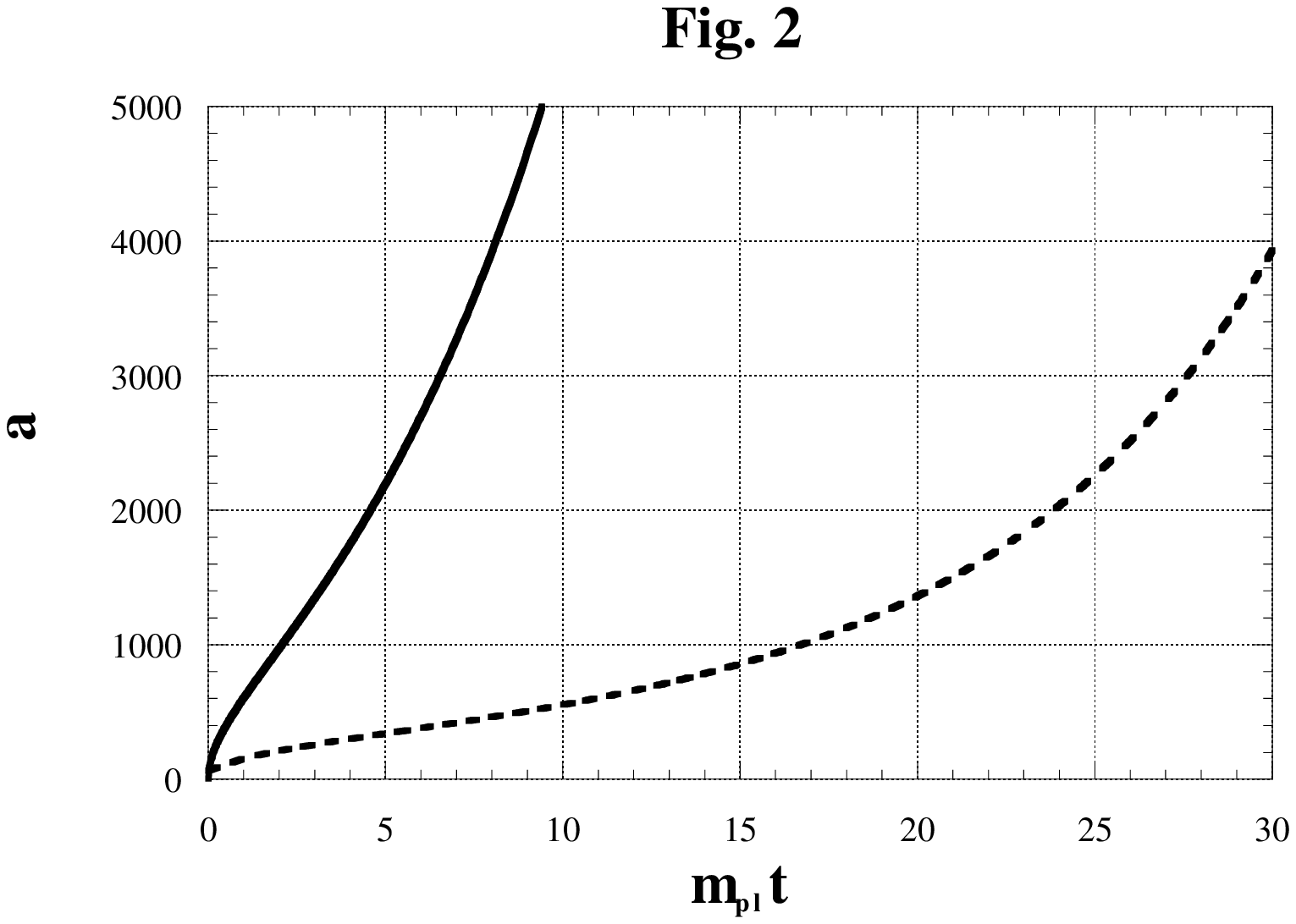}
\begin{figcaption}{Fig2}{12cm}
The evolution of the scale factor $a$ as a function of $t$
in the case of $\xi=0.05$ and $p=1/2$ with the initial values
of $\phi_i=-0.882m_{\rm pl}$ (solid) and   
$\phi_i=0.891m_{\rm pl}$ (dotted).
In both cases, we find that inflation begins to take place
after the non-inflationary evolution at the initial stage.
\end{figcaption}
\end{center}
\end{figure}

Let us next examine the case where $\phi$ is initially in the range
of $\phi_*<\phi_i<\phi_c$. 
If $\phi_i$ is close to the critical value $\phi_c$,
we have the non-inflationary solution $(\ref{C8})$ 
in any value of $p$ and $\xi$. 
However, as $\phi$ approaches the value of $\phi_*$, the assisted 
mechanism begins to work. 
Let us consider the case of $p=1/2$ and $\xi=0.05$ 
with the initial value of $\phi$ close to $\phi_c=0.892m_{\rm pl}$.
We find in Fig.~2 the inflationary behavior
after $\bar{t} \approx 5$, although the universe evolves 
deceleratedly as Eq.~$(\ref{C8})$ at the initial stage.
At $\bar{t}=5$, the value of inflaton is 
$\phi=0.880m_{\rm pl}$ (See Fig.~3), 
which is close to the critical value $\phi_c$.
This means that we have the non-inflationary solution $(\ref{C8})$
only when $\phi$ is very close to $\phi_c$.
Nonminimal coupling drives inflation while inflaton 
slowly evolves in the flat region
of $\phi_*<\phi~\mbox{\raisebox{-1.ex}{$\stackrel
     {\textstyle<}{\textstyle \sim}$}}~0.880m_{\rm pl}$. Finally, inflaton reaches
the local minimum of the effective potential, as is the same with 
the case of $-\phi_c<\phi_i<\phi_*$.

Next, we investigate the case where $p$ and $\xi$ are changed.
As is found by Eq.~$(\ref{C5})$, $\phi_*$ decreases
with the increase of $\xi$.
Since the effective potential $\hat{V}(\phi)$ 
becomes flatter in the region of
$0<\phi<\phi_*$ as $\xi$ increases, inflation is more 
easily realized 
in this region in spite of the decrease of $\phi_*$.
In the case of $\phi_*<\phi<\phi_c$, since 
the flat region around the 
local minimum becomes wider for larger value of $\xi$,
assisted inflation occurs significantly except the case where
$\phi$ is close to $\phi_c$.
This suggested that assisted mechanism works more efficiently 
with the increase of $\xi$ in the region of $\phi>0$,
and we have numerically confirmed this fact.

If we choose the values of $p$ with $p>1$,  inflation
is always supported by the effect of positive $\xi$ 
when inflaton evolves in the region of
$\phi>0$ as long as $\phi$ is not close to $\phi_c$.
On the other hand, since the slope of the effective potential 
becomes steeper with the decrease of $p$, we do not necessarily
have inflationary solutions for the case of $p \le 1$.
In the case where nonminimal coupling makes 
the effective potential flatter than in the case of 
$p=1$ and $\xi=0$, assisted inflation occurs
in the region of $\phi>0$.
For example, for $p=2/3$ and $p=1/2$ cases,
the coupling $\xi$ is required to be 
$\xi~\mbox{\raisebox{-1.ex}{$\stackrel
     {\textstyle>}{\textstyle\sim}$}}~3 \times 10^{-3}$
and $\xi~\mbox{\raisebox{-1.ex}{$\stackrel
     {\textstyle>}{\textstyle\sim}$}}~7 \times 10^{-3}$,
respectively, to lead to 
assisted inflation while $\phi$ rolls down toward $\phi_*$.
Although it is rather difficult to obtain the sufficient number of 
$e$-foldings to solve the cosmological puzzles only by the effect
of nonminimal coupling in the case of $p\le 1$, it may be 
possible to lead to sufficient inflation in the multi-field case.
What we emphasize is that inflation is realized even in the 
one-field case with $p \le 1$ in taking into account the effect
of nonminimal coupling. 

\begin{figure}
\begin{center}
\singlefig{12cm}{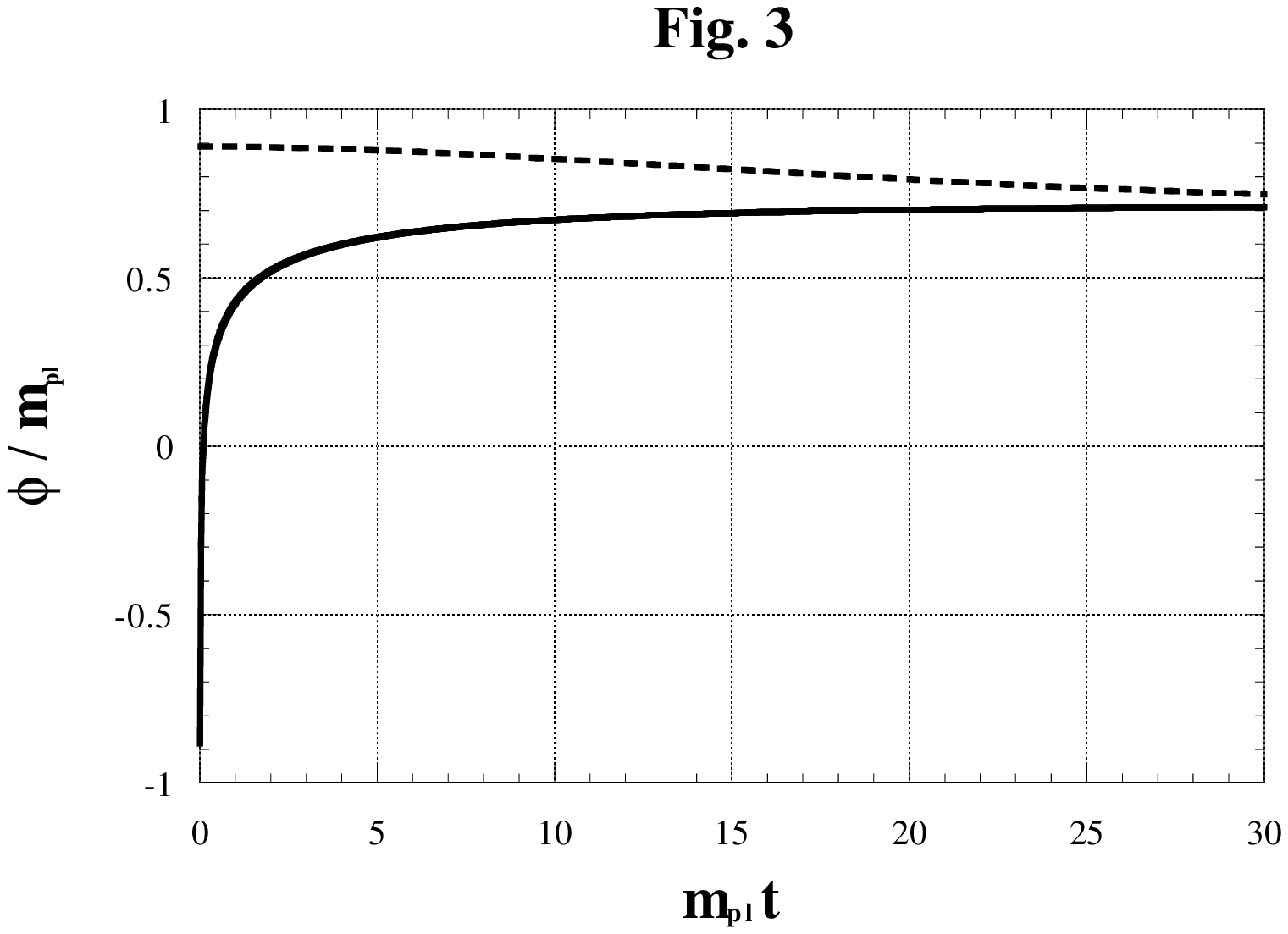}
\begin{figcaption}{Fig3}{12cm}
The evolution of the inflaton field $\phi$ as a function of $t$
in the case of $\xi=0.05$ and $p=1/2$ with the initial values
of $\phi_i=-0.882m_{\rm pl}$ (solid) and   
$\phi_i=0.891m_{\rm pl}$ (dotted).
In both cases, inflaton evolves toward the potential minimum
at $\phi_*=0.714m_{\rm pl}$.
\end{figcaption}
\end{center}
\end{figure}

\subsection{Case of $\xi<0$}   
When $\xi$ is negative, the shape of the effective 
potential $\hat{V}(\phi)$
is different depending on the relation of $\xi$ and $p$.
In the case of $|\xi|<1/2p$, the potential does not have any
extrema. However, when $|\xi|>1/2p$, 
it has both of a local minimum 
and a local maximum. In what follows,
we examine these two different cases separately.

\subsubsection{Case of $|\xi|<1/2p$}   
In this case, $\hat{V}(\phi)$ decreases monotonically with 
the increase of $\phi$. Although the potential is flatter than in 
the case of $\xi=0$ for negative values of $\phi$, it is steeper
for $\phi>0$ (See Fig.~1). 
Hence assisted inflation does not take place
when $\phi$ is initially located in the region of $\phi>0$.
However, for the initial values of $\phi<0$, we can expect
inflation to occur even in the case of $p \le 1$. 
For example, let us consider the case of $p=3/4$ with the initial value of 
$\phi_i=-2m_{\rm pl}$.
Numerical calculations show that we have an inflationary solution
at the initial stage for the coupling of 
$|\xi|~\mbox{\raisebox{-1.ex}{$\stackrel
     {\textstyle>}{\textstyle\sim}$}}~0.3$.
We depict in Fig.~4 the evolution of the scale factor $a$
as a function of $t$ for two cases of
$\xi=-0.5$ and $\xi=-0.2$. Although the universe evolves 
as non-inflationary at the whole stage
in the $\xi=-0.2$ case, inflation occurs 
in the $\xi=-0.5$ case at the initial stage where $\phi$
is negative. The larger values of $|\xi|$
($~\mbox{\raisebox{-1.ex}{$\stackrel
     {\textstyle>}{\textstyle\sim}$}}~0.3$) 
lead to the larger amount of inflation.  
If $\phi$ is initially  smaller than the value of 
$\phi_i=-2m_{\rm pl}$, one may consider that  
we will obtain the larger amount of 
$e$-foldings. However, this is not the case.
With the decrease of $\phi$, since the $-\xi\kappa^2\phi^2$
term in Eq.~$(\ref{B18})$ increases and becomes much larger
than unity, the velocity of the $\phi$ field is larger than that of
the $\Phi$ field. This suggests that the $\phi$ field does not 
evolve slowly enough to drive inflation for the smaller values 
of $\phi$.
For example, in the case of $p=3/4$ and $\xi=-0.5$,
we have numerically found that the universe evolves as
non-inflationary for the values of 
$\phi_i~\mbox{\raisebox{-1.ex}{$\stackrel
     {\textstyle<}{\textstyle \sim}$}}~-4m_{\rm pl}$.

When $p$ is less than unity, inflation is not realized
unless we choose rather large values of $|\xi|$.
For the case of $p=1/2$ with the initial value of 
$\phi_i=-2m_{\rm pl}$, we require the values of 
$|\xi|~\mbox{\raisebox{-1.ex}{$\stackrel
     {\textstyle>}{\textstyle\sim}$}}~0.6$ for inflation to occur.
However, we should stress that assisted inflation is possible
in the case of $p \le 1$ for the larger values of $|\xi|$
which satisfy $|\xi|<1/2p$. 

\begin{figure}
\begin{center}
\singlefig{12cm}{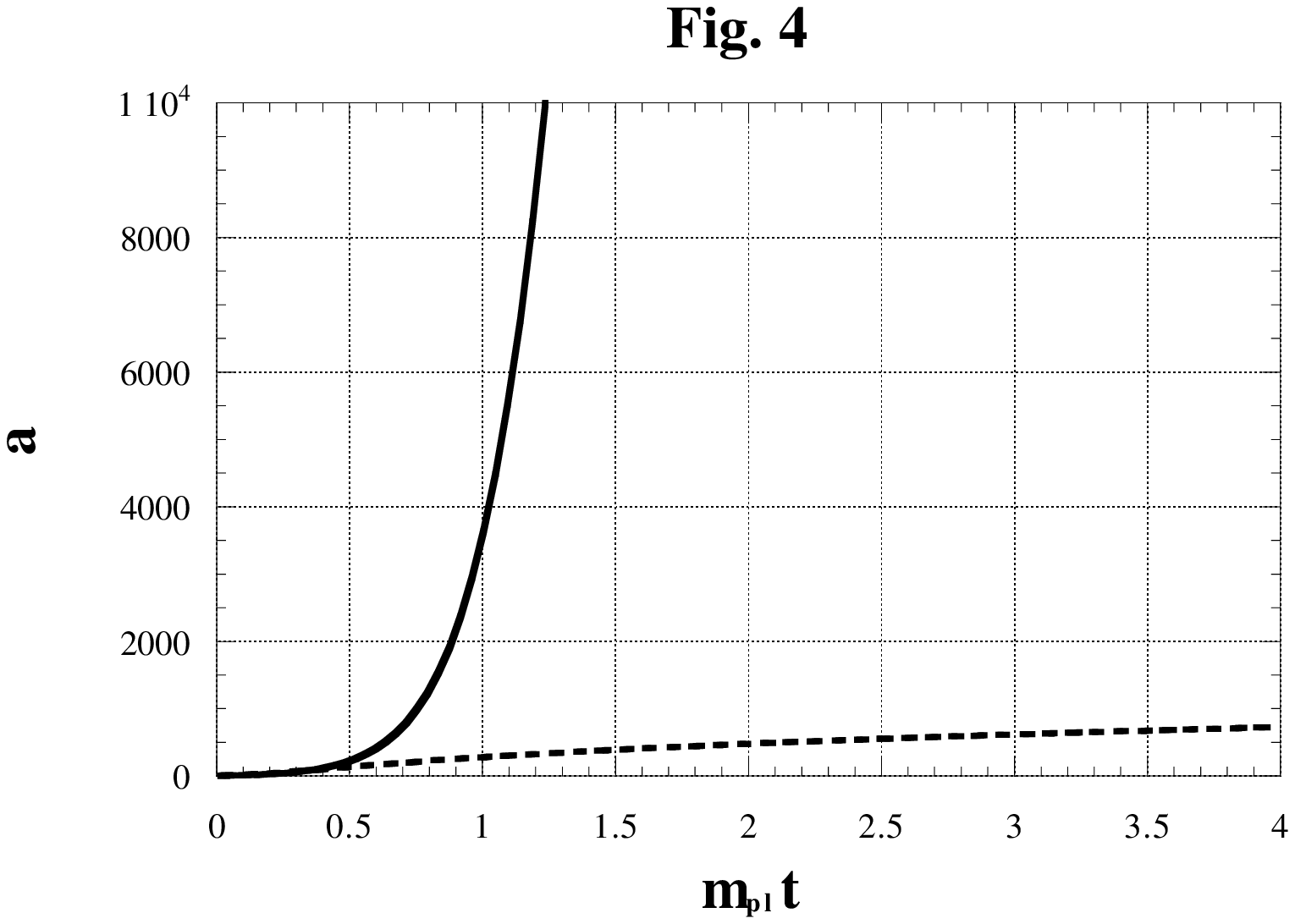}
\begin{figcaption}{Fig4}{12cm}
The evolution of the scale factor $a$ as a function of $t$
in the case of $\xi=-0.5$ (solid) and $\xi=-0.2$ (dotted)
with $p=3/4$ and the initial value of $\phi_i=-2m_{\rm pl}$.
In the case of $\xi=-0.5$, assisted inflation
by nonminimal coupling
relevantly occurs in the region of $\phi<0$.
On the other hand, when $\xi=-0.2$, 
inflation does not take place from the beginning.
\end{figcaption}
\end{center}
\end{figure}

\subsubsection{Case of $|\xi|>1/2p$}   
In this case, the potential $\hat{V}(\phi)$ is a local maximum at
\begin{eqnarray}
\phi_1=-\left( \sqrt{\frac{p}{4\pi}}-
\sqrt{\frac{p}{4\pi}+\frac{1}{8\pi\xi}} \right)
m_{\rm pl},
\label{C9}
\end{eqnarray}
and a local minimum at
\begin{eqnarray}
\phi_2=-\left( \sqrt{\frac{p}{4\pi}}+
\sqrt{\frac{p}{4\pi}+\frac{1}{8\pi\xi}} \right)
m_{\rm pl}.
\label{C10}
\end{eqnarray}
Note that both of $\phi_1$ and $\phi_2$ are negative.
With the existence of a local minimum, we can expect 
assisted inflation to occur in the region of $\phi<\phi_1$.
We consider the case of $p=1$ and $\xi=-1$ as one example.
In this case, $\phi_1$ and $\phi_2$ correspond to
$\phi_1=-0.083m_{\rm pl}$ and $\phi_2=-0.482m_{\rm pl}$,
respectively. When $\phi$ initially lies in $\phi_1<\phi_i<0$, 
assisted inflation occurs when inflaton
rolls down in the region of $\phi<0$ as 
in the case of $|\xi|<1/2p$. However, 
once inflaton passes through $\phi=0$, assisted mechanism 
does not work since the effective potential becomes steeper
compared with the $\xi=0$ case.
When the initial value of inflaton is less than $\phi_1$, the field evolves
toward the local minimum at $\phi=\phi_2$ and is finally trapped
in this minimum except the case where
$\phi_i$ is much smaller than $\phi_2$.
Inflation relevantly occurs as 
$\phi$ approaches the potential minimum, since
$\hat{V}(\phi)$ becomes gradually flatter.
We show in Fig.~5 the evolution of the scale factor  
in the original frame for two cases of 
$\phi_i=-0.090m_{\rm pl}$ and $\phi_i=-3m_{\rm pl}$.
In both cases, we find that the universe evolves as inflationary. 
Inflaton is finally trapped in the local minimum
at $\phi=\phi_2$ after $\bar{t} \approx 4$ for the 
$\phi_i=-0.090m_{\rm pl}$ case, and after $\bar{t} \approx 1.5$ 
for the $\phi_i=-3m_{\rm pl}$ case.
On the other hand, when the initial value of inflaton is
$\phi_i~\mbox{\raisebox{-1.ex}{$\stackrel
     {\textstyle<}{\textstyle \sim}$}}~-5m_{\rm pl}$,
the $\phi$ field moves rather rapidly and goes beyond
the local maximum at $\phi_1$. 
In this case, since the velocity of the $\phi$ field is large due to 
the relation of Eq.~$(\ref{B18})$, we have no
inflationary solution.  
Namely, in the case of $p=1$ and $\xi=-1$,
assisted inflation is realized  
for the initial values of 
$-5m_{\rm pl}~\mbox{\raisebox{-1.ex}{$\stackrel
     {\textstyle<}{\textstyle \sim}$}}~\phi_i<0$.

If $p$ is less than unity, we require rather large 
values of $|\xi|$ greater than the order of unity
due to the condition of $|\xi|>1/2p$.
However, it is important to note that nonminimal coupling
can lead to inflation even if $p \le 1$ in the region of $\phi<0$.
We have numerically found that assisted inflation can be 
realized for the values of $\xi$ which satisfy $|\xi|>1/2p$.

\begin{figure}
\begin{center}
\singlefig{12cm}{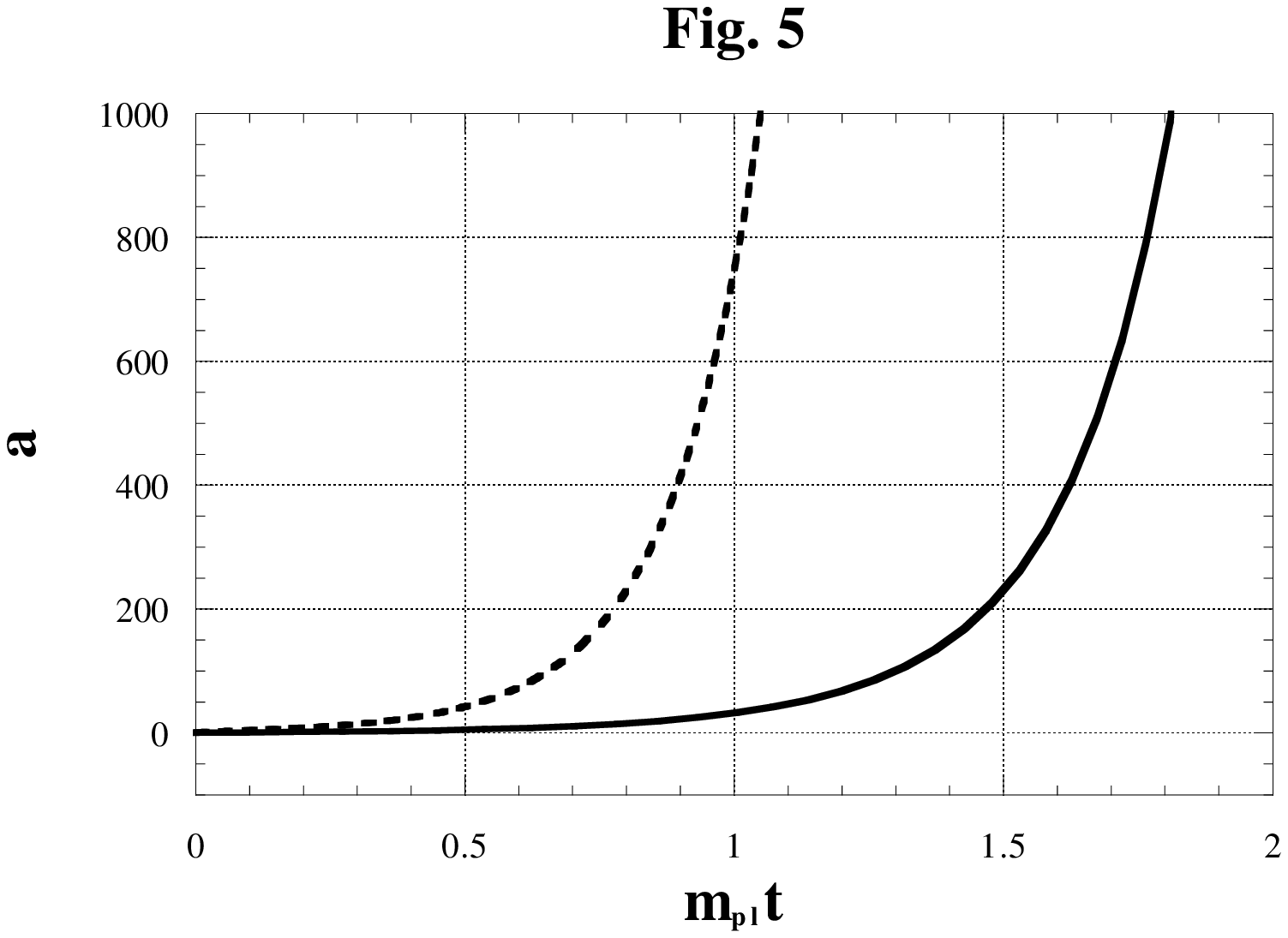}
\begin{figcaption}{Fig5}{12cm}
The evolution of the scale factor $a$ as a function of $t$
in the case of $\xi=-1$ and $p=1$ with the initial values of
$\phi_i=-0.090m_{\rm pl}$ (solid) and   
$\phi_i=-3m_{\rm pl}$ (dotted).
In both cases, inflation occurs while inflaton slowly
evolves toward the potential 
minimum at $\phi_2=-0.482m_{\rm pl}$.
\end{figcaption}
\end{center}
\end{figure}

\subsection{Summary}   
 Finally, we present two-dimensional plots of $\xi$ and
$p$ which divide the parameter regions of the inflationary 
and non-inflationary solutions in Fig.~6. 
These parameter spaces depend on  initial values of inflaton.
For $\phi_i \ge 0$, inflation can take 
place even when $p<1$ for positive values of $\xi$ (see Fig.~6a).
In this case, however, negative $\xi$ does not lead to assisted 
inflation because assisted mechanism is absent for $\phi>0$.
On the other hand, when inflaton is initially located for
$\phi_i<0$, inflationary behavior appears when inflaton evolves 
in the region of $\phi<0$ even for negative $\xi$ (see Fig.~6b).

\begin{figure}
\begin{center}
\singlefig{12cm}{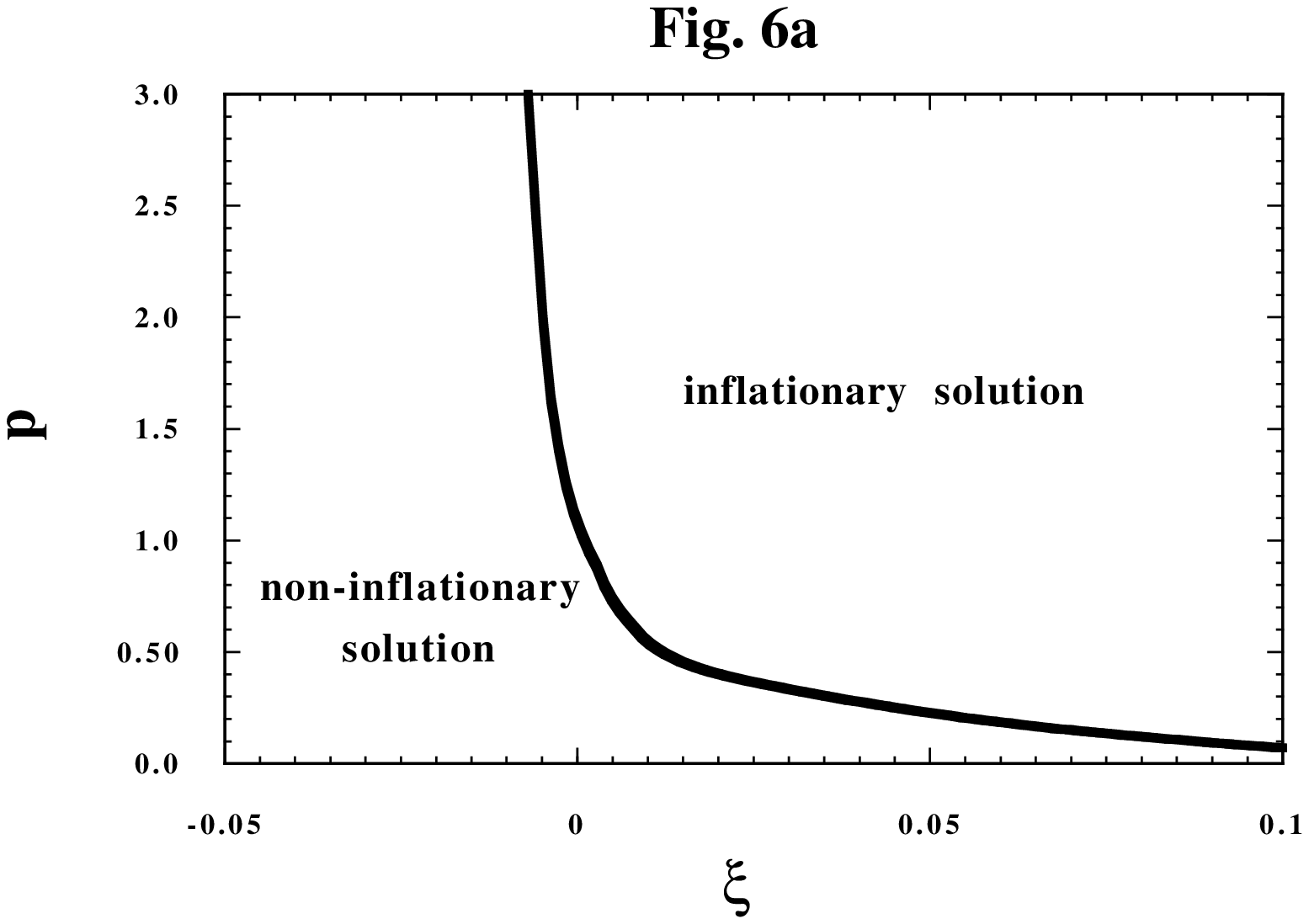}
\end{center}
\end{figure}
\begin{figure}
\begin{center}
\singlefig{12cm}{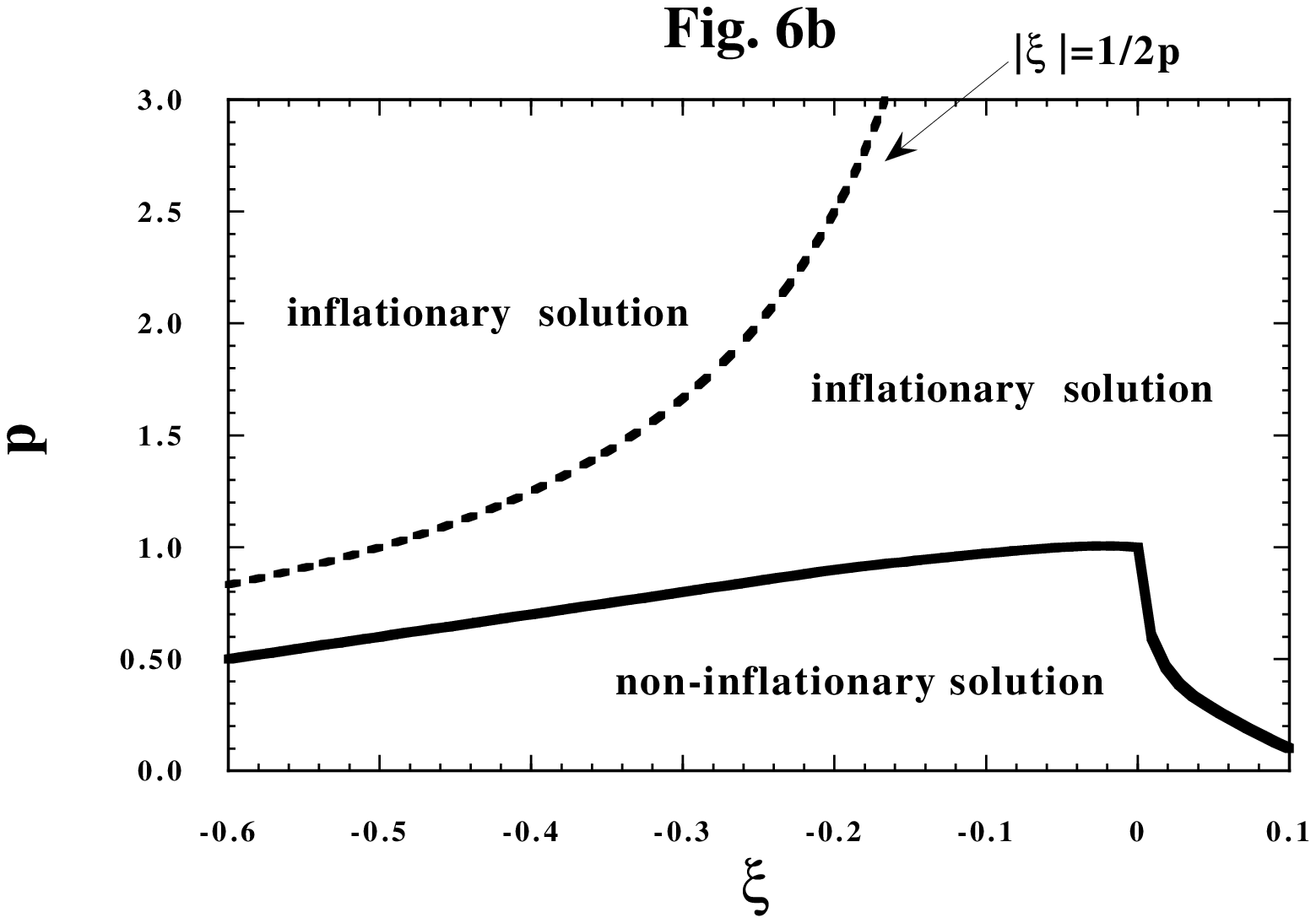}
\begin{figcaption}{Fig6b}{12cm}
The parameter regions of $\xi$ and $p$ where inflationary 
behavior appears for the initial values of $\phi_i=0$ (Fig.~6a)
and $\phi_i=-0.5m_{\rm pl}$ (Fig.~6b).
In both figures, the solid curves separate the regions of 
inflationary and non-inflationary solutions.
Assisted inflation can be realized for $\phi_i \ge 0$ in the case of
$\xi>0$. For $\phi_i <0$, negative $\xi$ also leads to assisted
inflation.
\end{figcaption}
\end{center}
\end{figure}

\section{Concluding remarks and discussions}   
In this paper, we have investigated the dynamics of 
power-law inflation with a nonminimally coupled inflaton field.
We studied how nonminimal coupling affects the 
dynamics of inflation with an exponential potential
$V(\phi)=V_0~{\rm exp}(-\sqrt{16\pi/p m_{\rm pl}^2} \phi)$
in the one-field model.

In the case of $\xi>0$, since an effective potential 
$(\ref{B11})$ in the Einstein frame which appears
by a conformal transformation becomes flatter
for $\phi>0$ than in the case of $\xi=0$, 
assisted inflation can be realized by nonminimal coupling.
In this case, nonminimal coupling 
gives rise to a potential minimum at some 
positive value of $\phi$,
and the potential becomes sufficiently flat around this area.
Assisted mechanism works except the case where
inflaton evolves in the region of $\phi<0$ and $\phi$ is close to 
the value of $\phi_c=1/\sqrt{8\pi \xi}$.
Even when  the power $p$ is less than unity, we have an 
inflationary solution  by choosing the appropriate 
values of $\xi$.
For example, we have numerically  found
that inflation occurs for 
$\xi~\mbox{\raisebox{-1.ex}{$\stackrel
     {\textstyle>}{\textstyle\sim}$}}~3 \times 10^{-3}$
when $p=2/3$, and for $\xi~\mbox{\raisebox{-1.ex}{$\stackrel
     {\textstyle>}{\textstyle\sim}$}}~7 \times 10^{-3}$ 
when $p=1/2$.

When $\xi$ is negative, the shape of the effective potential 
$\hat{V}(\phi)$ is different depending on two cases 
of $|\xi|<1/2p$ and $|\xi|>1/2p$. 
In the former case, $\hat{V}(\phi)$ is a monotonically decreasing
function of $\phi$. Since the potential is flat for $\phi<0$
compared with the case of $\xi=0$, assisted mechanism works
in this region as long as $\phi$ is not too far from $\phi=0$. 
However, nonminimal coupling prevents 
inflation for $\phi>0$.
In the latter case, the effective potential has a local maximum 
at $\phi_1$ and a local minimum at $\phi_2$ with 
$\phi_2<\phi_1<0$. When $\phi$ is initially in the range of 
$\phi_1<\phi<0$, the assisted dynamics is the same as
the former case.
When $\phi$ is smaller than $\phi_1$ initially,
inflaton evolves toward the 
potential minimum at $\phi_2$.
In this case, nonminimal coupling assists inflation to occur
unless $\phi$ is much smaller than $\phi_2$. 
For summary, we conclude that assisted 
inflation can be realized for $\phi>0$ in the case of 
$\xi>0$; and for $\phi<0$ in the case of 
$\xi<0$.
 
 The higher dimensional Kaluza-Klein 
 theories often give rise to exponential potentials 
which are obtained by means of a conformal transformation 
to the Einstein frame.
 Then there is a possibility that inflaton is minimally coupled
 to gravity in the effective four-dimensional theories.
 Although we did not ask the origin of  the exponential potential
 and considered a nonminimally coupled inflaton field
 in the four-dimensional action as in the 
 chaotic inflation model plus nonminimal coupling,
 we have to keep in mind that nonminimal coupling may not play
 relevant roles in realistic models of physics.
 What we emphasize is that assisted inflation is possible even 
 in the one-field model by introducing nonminimal coupling.
 When inflaton is minimally coupled to the spacetime curvature
 in the effective four-dimensional theories, inflation can be 
 assisted by considering multiple scalar fields. 
 
 The problem of power-law inflation with an 
 exponential potential is that the potential does not have
 a local minimum which causes a successful reheating.
 We find that introducing nonminimal coupling gives rise to   
 a local minimum to which inflaton rolls down.
 However, since inflaton continues to possess a constant
 energy at this minimum, the universe inflates forever.
 This ever-inflating problem also occurs in the minimally coupled
 case. One exit mechanism is to introduce another scalar field
 as the hybrid inflation model, by which inflaton is to evolve
 toward a true minimum. 
 Although we do not consider realistic models
  of exponential potentials which have 
 graceful exit from inflation in this paper, it is of interest
 whether nonminimal coupling or multiple scalar fields assist
 inflation or not in such models. 
 These issues are under consideration.

\section*{ACKOWLEDGEMENTS}
The author would like to thank Bruce A. Bassett, Kei-ichi Maeda,
Nobuyuki Sakai, Takashi Torii, and Kohta Yamamoto 
for useful discussions. This work was supported partially 
by a Grant-in-Aid for  Scientific
Research Fund of the Ministry of Education, Science and Culture
(No. 09410217) and by the Waseda University Grant 
for Special Research Projects.


\end{document}